# TreatmentEstimatoR: a Dashboard for Estimating Treatment Effects from Observational Health Data

Collin Sakal[1,2], Hon Hwang[1], Juan C Quiroz[1], and Blanca Gallego[1]*

**Affiliations:**

1.  Centre for Big Data Research in Health, University of New South Wales, Sydney, Australia

2.  School of Data Science, City University of Hong Kong, Hong Kong SAR, China

**\*Address correspondence to:** Blanca Gallego, b.gallego@unsw.edu.au

**Abstract**

*Motivation:* Observational health data can be leveraged to measure the real-world use and potential benefits or risks of existing medical interventions. However, lack of programming proficiency and advanced knowledge of causal inference methods excludes some clinicians and non-computational researchers from performing such analyses. Code-free dashboard tools provide accessible means to estimate and visualize treatment effects from observational health data.

*Results:* We present TreatmentEstimatoR, an R Shiny dashboard that facilitates the estimation of treatment effects from observational data without any programming knowledge required. The dashboard provides effect estimates from multiple algorithms simultaneously and accommodates binary, continuous, and time-to-event outcomes. TreatmentEstimatoR allows for flexible covariate selection for treatment and outcome models, comprehensive model performance metrics, and an exploratory data analysis tool.

*Availability:* TreatmentEstimatoR is available at https://github.com/CollinSakal/TreatmentEstimatoR. We provide full installation instructions and detailed vignettes for how to best use the dashboard.

# 1. Introduction

Estimating treatment effects from observational data involves adjusting for the differences in baseline characteristics between subjects in the treated and control groups, to compensate for the lack of treatment randomization. Many popular methods achieve this covariate balance between treatment and control groups by stratifying, matching, or weighing each subject according to their conditional probability of belonging to the treated group (i.e., the propensity score). Therefore, causal inference estimates often require: (1) an accurate estimate of the treatment model, (2) an accurate estimate of the outcome model, and (3) a valid strategy to adjust the outcome model according to the treatment mechanism [1]. Each of these steps is error-prone, either due to complex study designs or programmatic errors when writing the code to perform the analysis. In addition, visualization of the balance in covariate distributions between treated and control groups is important for the analyst to understand the validity and generalizability of the effect estimates.

Dashboard tools that expand the accessibility of statistical analyses are being popularized through development software such as R's Shiny package. Here, we present TreatmentEstimatoR, a Shiny dashboard that guides users through the steps to calculate treatment effects, automates the estimation of treatment effects using multiple statistical and machine learning algorithms, and provides the user with interactive data analysis and visualization tools.

# 2. Methods

TreatmentEstimatoR is freely available through GitHub at https://github.com/CollinSakal/TreatmentEstimatoR. The dashboard was developed in R using the Shiny and shinydashboard packages. TreatmentEstimatoR can handle three types of outcomes: binary, continuous, and time-to-event (survival). For binary and continuous outcome analyses, the dashboard includes Inverse Probability Weighting (IPW), a SuperLearner based Targeted Maximum Likelihood Estimator (TMLE), and Bayesian Additive Regression Trees (BART) [2]. For analysis of survival data, the dashboard includes Cox Proportional Hazard models with inverse probability weights, BART, and TMLE. TreatmentEstimatoR accepts comma-separated (csv) files and requires a dichotomous treatment variable. Data from the 2019 Atlantic Causal Inference Conference Data Challenge (including values of treatment effects) was used for testing TreatmentEstimatoR (see Supplementary Materials).

## 3. Software description

TreatmentExploreR is divided into three sections based on the type of analysis: binary, continuous, and survival. Each analysis type includes three subsections: Data Import, Summary Statistics, and Results. See Supplementary Materials for a full description of TreatmentEstimatoR.

**3.1 Data Import Section**

The data import page for binary and continuous outcome analyses prompts the user to upload a csv file containing column headers, an outcome column, a dichotomous treatment column, and at least one covariate column. Next, the user must specify information about the data: which columns represent the outcome and treatment, which covariates are categorical, which covariates should be removed from the treatment models or outcome models. For survival analysis, the user must specify the columns that correspond to observation start and end dates, and a cut-off value (follow-up) to calculate censoring. Lastly, a metric to estimate must be chosen from average treatment effect (ATE), average treatment effect on the treated, and average treatment effect on the controls [3].

### 3.2 Summary Statistics Section

The Summary Statistics page displays basic information about the data: the number of covariates and subjects, the percentage of subjects receiving treatment, the percentage experiencing the outcome, and what percent of the data are missing. An exploratory data analysis (EDA) tool enables users to obtain detailed information about a selected variable. For categorical variables, the EDA tool will return the category proportions and a distribution plot stratified by treatment assignment. For continuous variables the EDA tool displays summary statistics, a histogram with density overlay, and a distributional plot stratified by treatment assignment. The Summary Statistics page also automatically generates a baseline characteristics table (i.e., "Table 1"), which can be copied directly from the dashboard into an editor for further editing. Every plot in the Summary Statistics page can be saved directly as an image, enabling images to quickly be incorporated into manuscripts.

### 3.3 Results Section

For binary and continuous outcomes, the results page features a forest plot showing the treatment effect estimates from IPW, TMLE, and BART (Fig. 1). Point and interval estimates are provided along with p-values. For each algorithm, the dashboard displays a propensity score plot stratified by treatment assignment. For survival analysis, the results page calculates estimates using TMLE, BART, and a Cox Proportional Hazard Model. We defined ATE in the survival context as the difference in survival curves between the treated and untreated groups, with an interactive plot allowing users to examine ATE across all time points.

Metrics for evaluating the quality of the predictions of each algorithm's treatment and outcome models are calculated using 5-fold cross-validation. The area under the receiver operating characteristic curve is used for evaluating treatment models and binary outcome models, continuous outcome models are evaluated using mean squared error, and the concordance index is used for survival outcome models. For assessing calibration, the results page includes Brier scores and calibration plots.

## 4. Conclusion

Observational health data sources provide valuable opportunities for estimating treatment effects without some of the complications posed by conducting randomized clinical trials. Yet, programming proficiency and the complexity of causal inference methods remain sizable impediments for researchers wishing to analyze such data. We created a freely available dashboard, TreatmentEstimatoR, which streamlines the process of estimating treatment effects from observational health data without any programming required. This has the potential to lower the barrier to estimate treatment effects using causal inference methods, speed up the conduct of this research and corresponding publications, and improve the reproducibility of causal inference studies.

## Funding

This work was supported by the National Health and Medical Research Council, project grant No. APP1184304.

*Conflict of Interest:* none declared.

## References

1.) Vincent Dorie, Jennifer Hill, Uri Shalit, Marc Scott, Dan Cervone. Automated versus Do-It-Yourself Methods for Causal Inference: Lessons Learned from a Data Analysis Competition. Statist. Sci. 34 (1) 43 - 68, February 2019. https://doi.org/10.1214/18-STS667.

2.) Liuyi Yao, Zhixuan Chu, Sheng Li, Yaliang Li, Jing Gao, and Aidong Zhang. 2021. A Survey on Causal Inference. ACM Trans. Knowl. Discov. Data 15, 5, Article 74 (June 2021), 46 pages. https://doi-org/10.1145/3444944.

3.) Erin Hartman, Richard Grieve, Roland Ramsahai, Jasjeet Sekhon. From sample average treatment effect to population average treatment effect on the treated: Combining experimental with observational studies to estimate population treatment effects. Journal of the Royal Statistical Society: Series A (Statistics in Society), 178, 2015. https://doi-org/10.1111/rssa.12094.

Figure 1. Results screen. The results page shows the automatically calculated treatment effect estimates for multiple algorithms, a propensity score plot stratified by treatment assignment, and model performance statistics for treatment and outcome models

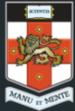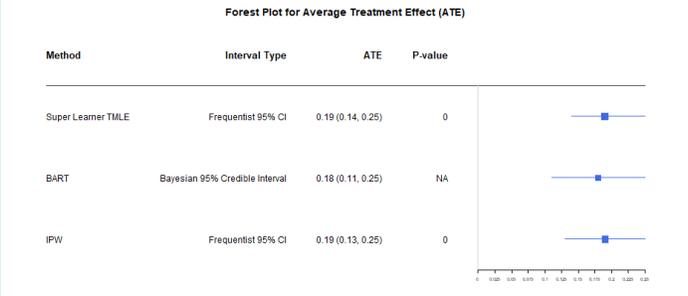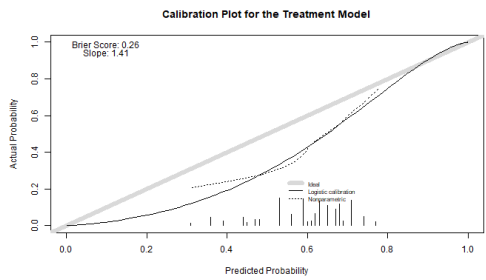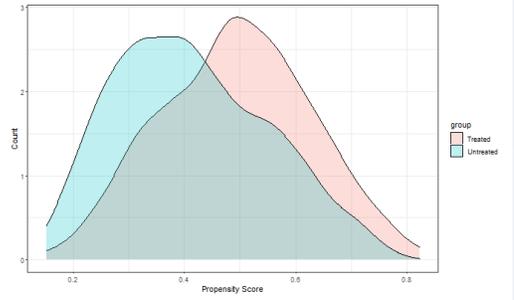

# Supplementary Materials

## 1. Installation Instructions

TreatmentEstimatoR is freely available at https://github.com/CollinSakal/TreatmentEstimatoR. Before being able to run TreatmentEstimatoR the local device must have R and RStudio downloaded. The first step is to download the repository as a ZIP file and store it in a local directory. Thereafter, navigate to RStudio and open the "server.R" or "ui.R" file and click "Run App" in the top right corner. R Studio may prompt you to download all of the necessary packages if they are not installed already, click yes in order to proceed.

## 2. Binary Data Analysis Walkthrough

Estimating treatment effects for binary outcome data begins with obtaining a properly formatted data set. TreatmentEstimatoR requires a comma separated (.csv) file with column headings, and in the binary case, a data set with dichotomous outcome and treatment columns, and at least one covariate column.

Next, open TreatmentEstimatoR and navigate to the "Binary Data Analysis" section and the "Data Import" submenu. Click browse and upload the data. Select which column name corresponds to the outcome column and which value within the outcome column corresponds to the outcome occurring. Repeat this process where indicated for the treatment column. Next, select which columns in the data are categorical. TreatmentEstimatoR will use all remaining columns (i.e. all columns that are not the outcome and treatment) as covariates in both the treatment model and the outcome model automatically. However, certain columns may be removed from either model in the "Remove Unwanted Variables" box. Lastly, select a metric to estimate from Average Treatment Effect (ATE), Average Treatment Effect on the Treated (ATT), and Average Treatment Effect on the Controls (ATC). Click "Start Analysis" to indicate to TreatmentEstimatoR that all the necessary data specifications have been made. If any errors occurred in the data import and specification process simply click "Reset Dashboard for New Analysis" in the Import Data box to start over.

Navigate to the "Summary Statistics" page or the "Results" page. We will describe the Summary Statistics page first. Upon opening there will be a box with summarizing information on the data: number of covariates, subjects, percentage of data missing, percentage of subjects who received treatment, and percentage who experienced the outcome. The "Summary Statistics About a Specific Variable" box will prompt the selection of a variable and whether it is categorical or not. For categorical variables category proportions and a barplot stratified by treatment will appear upon selection. For continuous variables TreatmentEstimatoR will return a histogram with density overlay, a density plot stratified by treatment, and a few statistics: minimum, mean, median, and maximum. Furthermore, there is a dropdown menu to select variables to include in a correlation matrix which will be generated in the right adjacent box. Lastly, there is a Table 1 creator, which automates the creation of Table 1s found in medical and epidemiological papers. This table can be copied and pasted directly into excel for editing.

The results page features a forest plot of treatment effect estimates from three algorithms: Inverse Probability Weighting (IPW), a SuperLearner based Targeted Maximum Likelihood Estimator (TMLE), and Bayesian Additive Regression Trees (BART). For more information on the algorithms please refer to the About the Algorithms section. Frequentist 95% confidence intervals, point estimates, and p-values are provided for IPW and TMLE. For BART a 95% Bayesian credible interval is provided along with a point estimate. Each treatment and outcome model is independently cross validated. Information on the area under the receiver operating characteristic curve (AUC) across the 5 cross-validation folds and a calibration plot can be found in the Method Performance Statistics box. Propensity score plots stratified by treatment assignment are also present in the Results Section.

Figures 1-4 show the data import, summary statistics, and results pages for the binary data analysis section.

## 3. Continuous Data Analysis Walkthrough

Estimating treatment effects for continuous outcome data begins with obtaining an appropriately formatted data set. TreatmentEstimatoR requires a comma separated (.csv) file with column headings, and in the continuous case, a data set with an outcome column, a dichotomous treatment column, and at least one covariate column.

Next, open TreatmentEstimatoR and navigate to the "Continuous Data Analysis" section and the "Data Import" submenu. Click browse and upload the data. Select which column name corresponds to the outcome column, then select the name of the treatment column and which value in the column corresponds to receiving treatment. Next, select which covariates in the data are categorical. TreatmentEstimatoR will use all remaining columns (i.e. all columns that are not the outcome and treatment) as covariates in both the treatment model and the outcome model automatically. However, certain columns may be excluded from either model in the "Remove Unwanted Variables" box. Lastly, select a metric to estimate from Average Treatment Effect

(ATE), Average Treatment Effect on the Treated (ATT), and Average Treatment Effect on the Controls (ATC). Click "Start Analysis" to indicate to TreatmentEstimatoR that all the necessary data specifications have been made. If any errors occurred in the data import and specification process simply click "Reset Dashboard for New Analysis" in the Import Data box to start over.

Navigate to the "Summary Statistics" page or the "Results" page. We will describe the Summary Statistics page first. Upon opening there will be a box with summarizing information on the data: number of covariates, subjects, percentage of data missing, percentage of subjects who received treatment, and the mean outcome. The "Summary Statistics About a Specific Variable" box will prompt the selection of a variable and whether it is categorical or not. For categorical variables category proportions and a barplot stratified by treatment will appear upon selection. For continuous variables TreatmentEstimatoR will return a histogram with density overlay, a density plot stratified by treatment, and a few statistics: minimum, mean, median, and maximum. Furthermore, there is a dropdown menu to select variables to include in a correlation matrix which will be generated in the right adjacent box. Lastly, there is a Table 1 creator, which automates the creation of Table 1s found in medical and epidemiological papers. This table can be copied and pasted directly into excel for editing.

The results page features a forest plot of treatment effect estimates from three algorithms: Inverse Probability Weighting (IPW), a SuperLearner based Targeted Maximum Likelihood Estimator (TMLE), and Bayesian Additive Regression Trees (BART). For more information on the algorithms please refer to the About the Algorithms section. Frequentist 95% confidence intervals, point estimates, and p-values are provided for IPW and TMLE. For BART a 95% Bayesian credible interval is provided along with a point estimate. Each treatment and outcome model is independently cross validated. For the treatment model information on the area under the receiver operating characteristic curve (AUC) across the 5 cross-validation folds and a calibration plot can be found in the Method Performance Statistics box. For the outcome model information on the mean squares error (MSE) across the cross-validation folds can also be found. Propensity score plots stratified by treatment assignment are also present in the Results Section.

Figures 5-8 show the data import, summary statistics, and results pages for the continuous data analysis section.

## 4. Survival Data Analysis Walkthrough

Estimating treatment effects using TreatmentEstimatoR for survival begins with acquiring an appropriately formatted data set. TreatmentEstimatoR requires a comma separated (.csv) file with a dichotomous treatment column, a dichotomous event column, a person-level start date column, person-level end date column, and at least one covariate column. The start date column must be populated with subject level dates indicating when they began the study. The end date column must contain similar information for the end of the study or the last follow up time.

Next, open TreatmentEstimatoR and navigate to the "Survival Data Analysis" section and the "Data Import" submenu. Click browse and upload the data. Select which column name corresponds to the event column, and which value within the event column corresponds to the outcome occurring. Repeat this process where indicated for the treatment column. Next, select which columns in the data are categorical. TreatmentEstimatoR will use all columns that are not the event, treatment, start date, or end date as covariates in the treatment and survival models. However, additional columns may be excluded from either model in the "Remove Columns" dropdown menus. Next, indicate the start and end date columns, the date format, time units, and the time cutoff (in the units selected) for the study. Any subjects who did not have an event before the cutoff will be censored. Click "Start Analysis" to conclude the requisite actions in the Data Import Section.

Navigate to the "Summary Statistics" page or the "Results" page. We will describe the Summary Statistics page first. Upon opening there will be a box with summarizing information on the data: number of covariates, subjects, percentage of data missing, percentage of subjects who received treatment, percentage who experienced the event, percentage censored, and the average time to event. There is also a follow-up time plot stratified by censor status. The "Summary Statistics About a Specific Variable" box will prompt the selection of a variable and whether it is categorical or not. For categorical variables category proportions and a barplot stratified by treatment will appear upon selection. For continuous variables TreatmentEstimatoR will return a histogram with density overlay, a density plot stratified by treatment, and a few statistics: minimum, mean, median, and maximum. Furthermore, there is a dropdown menu to select variables to include in a correlation matrix which will be generated in the right adjacent box. Lastly, there is a Table 1 creator, which automates the creation of Table 1s found in medical and epidemiological papers. This table can be copied and pasted directly into excel for editing.

The results section features interactive survival and ATE plots for each algorithm. The ATE plot is simply the difference in survival curves with confidence estimates. Points of the curves for both the survival and ATE plots can be viewed at any time point by hovering the mouse over a specific point. Additionally, a Kaplan-Meier plot is shown along with propensity score plots for Cox PH and TMLE (the survival algorithm for BART does not incorporate propensity scores). Lastly, performance metrics are provided for the treatment and survival models. The treatment model metrics include the mean, minimum, maximum, and standard deviation for the area under the receiver operating characteristic curve (AUC) across five cross-validation folds; a calibration plot is also shown. For the survival models fivefold cross validation was also performed, but the metric shown is the C-index. Users should note that the TMLE and BART in the survival section are very computationally intensive, especially for cross-validation. We recommend only utilizing these algorithms if time or computing power are in abundance.

Figures 9-13 show the data import, summary statistics, and results pages for the survival data analysis sections.

## 5. Back End

TreatmentEstimatoR is built in R shiny and has a front-end user interface (UI) and a back end (server file). In short, the user interacts with the UI which, depending on the user's action, triggers backend functions that relay new information to the UI. The walkthroughs presented above showed how the user interacts with the UI. For completeness we now describe the back end for each outcome type: binary, continuous, and survival.

The binary and continuous outcome backends are essentially identical and will be described in parallel. Once the user has uploaded their data a series of filtering functions return three data sets: one for the summary statistics page, one for the treatment models, and one for the outcome models. This ensures the user can explore their whole data frame in the summary statistics tab even if they removed covariates from the treatment or outcome models. Moreover, the separate data sets for the treatment and outcome models allows for different covariates to be included in each. An outline of the backend is shown in figure 14.

The survival backend has a similar structure but with more functions to accommodate the additional output. After the user has uploaded their data and made the requisite specifications in the Data Import tab TreatmentEstimatoR creates three data frames: one for the summary statistics tab, one for the treatment models, and one for the survival models. This allows different covariates to be included in both models, while a complete view of the data can be observed in the summary statistics section. An outline of the backend is shown in figure 15.

## 6. Validation

In order to test the validity of TreatmentEstimatoR we used it to analyze data with known values of ATE. For the binary and continuous case we used data from the 2019 Atlantic Causal Inference Conference (ACIC) [1]. The survival analysis section is still under development and has yet to be tested. The type of data is described on the website as:

"Covariates were drawn from publicly available data and also simulated. Identifiability of the parameter is guaranteed, however challenges to estimation have been built-in to the processes for generating the binary treatment assignment, and binary or continuous outcome. These include non-linearity of the response surface, treatment effect heterogeneity, varying proportion of true confounders among the observed covariates, and near violations of the positivity assumption."

For the binary section we used the "low1" data set, which has a true ATE of 0.11. In the data import tab we did not indicate that any columns were categorical, and included all possible covariates in both the treatment and outcome models. These specifications were mainly due to the fact that we had no knowledge of what the covariates represented. The TreatmentEstimatoR estimates for ATE were as follows: 0.10 (0.01, 0.19) for BART, 0.12 (0.04, 0.19) for TMLE, and 0.12 (0.05, 0.2) for IPW. All interval estimates included the true ATE and all point estimates were fairly close to the true value.

For the continuous section we used the "low10" data set, which has a known ATE of -0.8. We again did not indicate that any columns were categorical, and included all available covariates in both the treatment and outcome models. The TreatmentEstimatoR estimates were -0.79 (-1, -0.58) for BART, -0.8 (-0.98, -0.62) for TMLE, and -0.83 (-1.08, -0.59) for IPW. Again all interval estimates included the true ATE and all point estimates were fairly close to the true value with BART estimating the true value exactly.

## 7. About the Algorithms

For each data type treatment effect estimates are given by three algorithms. For the binary and continuous outcomes these are: a SuperLearner based Targeted Maximum Likelihood Estimation approach (TMLE), Bayesian Additive Regression Trees (BART), and Inverse Probability Weighting (IPW) [2-8]. For survival data the algorithms are Cox Proportional Hazard models (Cox PH), One-shot TMLE, and BART [9-12]. Further descriptions of each algorithm can be found below

Inverse Probability Weighting (IPW) for the binary and continuous outcome sections first estimates the propensity score via a logistic regression. Thereafter it computes the inverse probability weights, or the inverse probability of being assigned treatment, and assigns them to each subject. The mean difference in outcome between the treated and untreated groups is computed to obtain treatment effect estimates. The IPW functions were largely based off of the *causalweight* package [13].

The SuperLearner based Targeted Maximum Likelihood procedure for the binary and continuous data types is a doubly robust estimator. It first uses a machine learning ensemble consisting of XGBoost, Multivariate Additive Regression Splines (MARS), and a lasso regularized regression to estimate the treatment mechanism and assign weights. Then, the outcome is estimated using an ensemble with the same algorithms. Finally, a TMLE adjustment is made to obtain effect estimates. This procedure was implemented using the SuperLearner [3] and TMLE [4] packages. For survival analysis we implemented a TMLE procedure that performs the adjustment at multiple time points from the MOSS library. [12]

The Bayesian Additive Regression Trees (BART) was implemented for the binary and continuous outcome sections using the bartCause package [8]. BART is a black box sum of trees model implemented under a Bayesian framework. The survival BART was implemented using the BART package [11].

The survival section also featured a Cox Proportional Hazard (Cox PH) model with inverse probability weights. The weights were calculated from propensity scores estimated by logistic regression. The Cox PH model was implemented using the *survival* package [10].

# 8. R Packages

A list of R packages used in the dashboard's main functions is presented below divided into a few sections: application framework, modeling, data filtering, data visualizations, and miscellaneous. For each package a brief note on how it was used is provided. Note that the tables do not include every R package used. The full citations of every R package can be found in the R Package Citations section.

**Application Framework**

| R Package and Version | Use |
|---|---|
| 1. shiny 1.6.0<br>2. shinydashboard 0.7.1<br>3. shinyjs 2.0.0<br>4. rJava 0.9-13 | 1. User Interface<br>2. User Interface<br>3. Data import pages<br>4. Required to build Shiny apps |

**Modelling**

| R Package and Version | Use |
|---|---|
| 1. tmle 1.5.0-1<br>2. MOSS 1.2.0<br>3. survtmle 1.1.1<br>4. bartCause 1.0-4<br>5. BART 2.9<br>6. bartMachine 1.2.6<br>7. bartMachineJARs 1.1<br>8. SuperLearner 2.0-28<br>9. glmnet 4.1-2<br>10. earth 5.3.0<br>11. survival 3.2-10 | 1. Binary and continuous outcome TMLE<br>2. Survival TMLE<br>3. Survival TMLE (required to run MOSS)<br>4. Binary and continuous BART<br>5. Survival BART<br>6. Cross validating binary and continuous BART<br>7. Required for bartMachine<br>8. Propensity score estimation, cross validation of treatment models<br>9. Used in SuperLearner ensembles for treatment and outcome estimation<br>10. Used in SuperLearner ensembles for treatment and outcome estimation<br>11. Cox PH model |

**Data Filtering**

| R Package and Version | Use |
|---|---|
| 1. dplyr 1.0.5<br>2. lubridate 1.7.10<br>3. tidyr 1.1.3<br>4. data.table 1.14.0<br>5. cobalt 4.3.1<br>6. WeightIt 0.12.0<br>7. caret 6.0-88 | 1. Used throughout the backend for any tasks requiring data manipulation<br>2. Used for manipulating dates in the survival data filtering functions<br>3. Used extensively for various data filtering tasks<br>4. For creating data tables<br>5. Used in conjunction with other packages for weighting in the Cox PH survival section<br>6. Used for weighting in the Cox PH models<br>7. Creating cross-validation folds |

**Data Visualizations**

| R Package and Version | Use |
|---|---|
| 1. ggplot2 3.3.5<br>2. tableone 0.12.0<br>3. forestplot 1.10.1<br>4. rms 6.1-1<br>5. corrplot 0.89 | 1. All histograms, density plots, and bar plots<br>2. Table 1 generator in the summary statistics pages<br>3. Forest plots in the binary and continuous results page<br>4. Calibration plots for treatment model cross validation and binary outcome model<br>5. Correlation plots in the summary statistics tabs |

# References

1. https://sites.google.com/view/acic2019datachallenge/home
2. Austin PC. An Introduction to Propensity Score Methods for Reducing the Effects of Confounding in Observational Studies. Multivariate Behav Res. 2011;46(3):399-424. doi:10.1080/00273171.2011.568786
3. Eric Polley, Erin LeDell, Chris Kennedy and Mark van der Laan (2021). SuperLearner: Super Learner Prediction. R package version 2.0-28. https://CRAN.R-project.org/package=SuperLearner
4. Susan Gruber, Mark J. van der Laan (2012). tmle: An R Package for Targeted Maximum Likelihood Estimation. Journal of Statistical Software, 51(13), 1-35. URL http://www.jstatsoft.org/v51/i13/.
5. Megan S. Schuler, Sherri Rose, Targeted Maximum Likelihood Estimation for Causal Inference in Observational Studies, *American Journal of Epidemiology*, Volume 185, Issue 1, 1 January 2017, Pages 65–73, https://doi.org/10.1093/aje/kww165
6. van der Laan, Mark J. and Rubin, Daniel. "Targeted Maximum Likelihood Learning" *The International Journal of Biostatistics*, vol. 2, no. 1, 2006. https://doi.org/10.2202/1557-4679.1043
7. Hugh A. Chipman. Edward I. George. Robert E. McCulloch. "BART: Bayesian additive regression trees." Ann. Appl. Stat. 4 (1) 266 - 298, March 2010. https://doi.org/10.1214/09-AOAS285
8. Hill J (2011). "Bayesian Nonparametric Modeling for Causal Inference." _Journal of Computational and Graphical Statistics_, *20*(1), 217-240. doi: 10.1198/jcgs.2010.08162 (URL: https://doi.org/10.1198/jcgs.2010.08162).

9. Buchanan AL, Hudgens MG, Cole SR, Lau B, Adimora AA; Women's Interagency HIV Study. Worth the weight: using inverse probability weighted Cox models in AIDS research. *AIDS Res Hum Retroviruses*. 2014;30(12):1170-1177. doi:10.1089/AID.2014.0037

10. Therneau T (2021). _A Package for Survival Analysis in R_. R package version 3.2-10, URL: https://CRAN.R-project.org/package=survival.

11. Sparapani R, Spanbauer C, McCulloch R (2021). "Nonparametric Machine Learning and Efficient Computation with Bayesian Additive Regression Trees: The BART R Package." _Journal of Statistical Software_, *97*(1), 1-66. doi: 10.18637/jss.v097.i01 (URL: https://doi.org/10.18637/jss.v097.i01).

12. Cai W, van der Laan MJ (2019+). *One-step TMLE for time-to-event outcomes.* Working paper.

13. Hugo Bodory and Martin Huber (2020). causalweight: Estimation Methods for Causal Inference Based on Inverse Probability Weighting. R package version 1.0.0. https://CRAN.R-project.org/package=causalweight

## R Package Citations

1. R Core Team (2021). R: A language and environment for statistical computing. R Foundation for Statistical Computing, Vienna, Austria. URL https://www.R-project.org/.

2. Weixin Cai and Mark van der Laan (2021). MOSS: One-Step TMLE for Survival Analysis. R package version 1.2.0. https://github.com/wilsoncai1992/MOSS

3. Alboukadel Kassambara (2020). ggpubr: 'ggplot2' Based Publication Ready Plots. R package version 0.4.0. https://CRAN.R-project.org/package=ggpubr

4. Frank E Harrell Jr (2021). rms: Regression Modeling Strategies. R package version 6.1-1. https://CRAN.R-project.org/package=rms

5. Daniel J. Stekhoven (2013). missForest: Nonparametric Missing Value Imputation using Random Forest. R package version 1.4.

6. Simon Urbanek (2020). rJava: Low-Level R to Java Interface. R package version 0.9-13. https://CRAN.R-project.org/package=rJava

7. Stephen Milborrow (2020). plotmo: Plot a Model's Residuals, Response, and Partial Dependence Plots. R package version 3.6.0. https://CRAN.R-project.org/package=plotmo

8. Lang M (2017). "checkmate: Fast Argument Checks for Defensive R Programming." _The R Journal_, *9*(1), 437-445. <URL:https://journal.r-project.org/archive/2017/RJ-2017-028/index.html>.

9. Alexandros Karatzoglou, Alex Smola, Kurt Hornik, Achim Zeileis (2004). kernlab - An S4 Package for Kernel Methods in R. Journal of Statistical Software 11(9), 1-20. URL http://www.jstatsoft.org/v11/i09/

10. Dean Attali (2020). shinyjs: Easily Improve the User Experience of Your Shiny Apps in Seconds. R package version 2.0.0. https://CRAN.R-project.org/package=shinyjs

11. Douglas Bates and Martin Maechler (2021). Matrix: Sparse and Dense Matrix Classes and Methods. R package version 1.3-2. https://CRAN.R-project.org/package=Matrix

12. Hugo Bodory and Martin Huber (2020). causalweight: Estimation Methods for Causal Inference Based on Inverse Probability Weighting. R  package version 1.0.0. https://CRAN.R-project.org/package=causalweight

13. Winston Chang (2020). R6: Encapsulated Classes with Reference Semantics. R package version 2.5.0. https://CRAN.R-project.org/package=R6

14. Sparapani R, Spanbauer C, McCulloch R (2021). "Nonparametric Machine Learning and Efficient Computation with Bayesian Additive Regression Trees: The BART R Package." _Journal of Statistical Software_, *97*(1), 1-66. doi: 10.18637/jss.v097.i01 (URL: https://doi.org/10.18637/jss.v097.i01).

15. Roger Koenker and Pin Ng (2019). SparseM: Sparse Linear Algebra. R package version 1.78. https://CRAN.R-project.org/package=SparseM

16. Steve Weston and Hadley Wickham (2014). itertools: Iterator Tools. R package version 0.1-3. https://CRAN.R-project.org/package=itertools

17. Xavier Robin, Natacha Turck, Alexandre Hainard, Natalia Tiberti, Frédérique Lisacek, Jean-Charles Sanchez and Markus Müller (2011). pROC: an open-source package for R and S+ to analyze and compare ROC curves. BMC Bioinformatics, 12, p. 77. DOI: 10.1186/1471-2105-12-77 <http://www.biomedcentral.com/1471-2105/12/77/>

18. Greg Snow (2020). TeachingDemos: Demonstrations for Teaching and Learning. R package version 2.12. https://CRAN.R-project.org/package=TeachingDemos

19. Stefan Milton Bache and Hadley Wickham (2020). magrittr: A Forward-Pipe Operator for R. R package version 2.0.1. https://CRAN.R-project.org/package=magrittr

20. H. Wickham. ggplot2: Elegant Graphics for Data Analysis. Springer-Verlag New York, 2016.

21. Winston Chang and Barbara Borges Ribeiro (2018). shinydashboard: Create Dashboards with 'Shiny'. R package version 0.7.1. https://CRAN.R-project.org/package=shinydashboard

22. Eric Polley, Erin LeDell, Chris Kennedy and Mark van der Laan (2021). SuperLearner: Super Learner Prediction. R package version 2.0-28. https://CRAN.R-project.org/package=SuperLearner

23. Marvin N. Wright, Andreas Ziegler (2017). ranger: A Fast Implementation of Random Forests for High Dimensional Data in C++ and R. Journal of Statistical Software, 77(1), 1-17. doi:10.18637/jss.v077.i01

24. David Benkeser and Nima Hejazi (2019). survtmle: Compute Targeted Minimum Loss-Based Estimates in Right-Censored Survival Settings. R package version 1.1.1. https://CRAN.R-project.org/package=survtmle

25. Venables, W. N. & Ripley, B. D. (2002) Modern Applied Statistics with S. Fourth Edition. Springer, New York. ISBN 0-387-95457-0

26. Frank E Harrell Jr, with contributions from Charles Dupont and many others. (2021). Hmisc: Harrell Miscellaneous. R package version 4.5-0. https://CRAN.R-project.org/package=Hmisc

27. Revolution Analytics and Steve Weston (2020). iterators: Provides Iterator Construct. R package version 1.0.13. https://CRAN.R-project.org/package=iterators

28. Max Kuhn (2021). caret: Classification and Regression Training. R package version 6.0-88. https://CRAN.R-project.org/package=caret

29. Lemon, J. (2006) Plotrix: a package in the red light district of R. R-News, 6(4): 8-12.

30. Daniel E. Ho, Kosuke Imai, Gary King, Elizabeth A. Stuart (2011). MatchIt: Nonparametric Preprocessing for Parametric Causal Inference. Journal of Statistical Software, Vol. 42, No. 8, pp. 1-28. URL https://www.jstatsoft.org/v42/i08/

31. Hill J (2011). "Bayesian Nonparametric Modeling for Causal Inference." _Journal of Computational and Graphical Statistics_, *20*(1), 217-240. doi: 10.1198/jcgs.2010.08162 (URL: https://doi.org/10.1198/jcgs.2010.08162).

32. Winston Chang, Joe Cheng, JJ Allaire, Carson Sievert, Barret Schloerke, Yihui Xie, Jeff Allen, Jonathan McPherson, Alan Dipert and Barbara Borges (2021). shiny: Web Application Framework for R. R package version 1.6.0. https://CRAN.R-project.org/package=shiny

33. Trevor Hastie (2020). gam: Generalized Additive Models. R package version 1.20. https://CRAN.R-project.org/package=gam

34. dam Kapelner, Justin Bleich and see COPYRIGHTS file for the authors of the java libraries (2018). bartMachineJARs: bartMachine JARs. R package version 1.1. https://CRAN.R-project.org/package=bartMachineJARs

35. Kazuki Yoshida and Alexander Bartel (2020). tableone: Create 'Table 1' to Describe Baseline Characteristics with or without Propensity Score Weights. R package version 0.12.0. https://CRAN.R-project.org/package=tableone

36. Pinheiro J, Bates D, DebRoy S, Sarkar D, R Core Team (2021). _nlme: Linear and Nonlinear Mixed Effects Models_. R package version 3.1-152, <URL: https://CRAN.R-project.org/package=nlme>.

37. Therneau T (2021). _A Package for Survival Analysis in R_. R package version 3.2-10, <URL: https://CRAN.R-project.org/package=survival>.

38. A. Liaw and M. Wiener (2002). Classification and Regression by randomForest. R News 2(3), 18--22.

39. Sarkar, Deepayan (2008) Lattice: Multivariate Data Visualization with R. Springer, New York. ISBN 978-0-387-75968-5

40. Zeileis A, Croissant Y (2010). "Extended Model Formulas in R: Multiple Parts and Multiple Responses." _Journal of Statistical Software_, *34*(1), 1-13. doi: 10.18637/jss.v034.i01 (URL: https://doi.org/10.18637/jss.v034.i01).

41. Noah Greifer (2021). WeightIt: Weighting for Covariate Balance in Observational Studies. R package version 0.12.0. https://CRAN.R-project.org/package=WeightIt

42. Hadley Wickham (2021). tidyr: Tidy Messy Data. R package version 1.1.3. https://CRAN.R-project.org/package=tidyr

43. Susan Gruber, Mark J. van der Laan (2012). tmle: An R Package for Targeted Maximum Likelihood Estimation. Journal of Statistical Software, 51(13), 1-35. URL <http://www.jstatsoft.org/v51/i13/>.

44. Microsoft and Steve Weston (2020). foreach: Provides Foreach Looping Construct. R package version 1.5.1. https://CRAN.R-project.org/package=foreach

45. Alboukadel Kassambara, Marcin Kosinski and Przemyslaw Biecek (2021). survminer: Drawing Survival Curves using 'ggplot2'. R package version 0.4.9. https://CRAN.R-project.org/package=survminer

46. Garrett Grolemund, Hadley Wickham (2011). Dates and Times Made Easy with lubridate. Journal of Statistical Software, 40(3), 1-25. URL https://www.jstatsoft.org/v40/i03/.

47. Stephen Milborrow. Derived from mda:mars by Trevor Hastie and Rob Tibshirani. Uses Alan Miller's Fortran utilities with Thomas Lumley's leaps wrapper. (2020). earth: Multivariate Adaptive Regression Splines. R package version 5.3.0. https://CRAN.R-project.org/package=earth

48. Max Gordon and Thomas Lumley (2020). forestplot: Advanced Forest Plot Using 'grid' Graphics. R package version 1.10.1. https://CRAN.R-project.org/package=forestplot

49. Noah Greifer (2021). cobalt: Covariate Balance Tables and Plots. R package version 4.3.1. https://CRAN.R-project.org/package=cobalt

50. Hadley Wickham, Romain François, Lionel Henry and Kirill Müller (2021). dplyr: A Grammar of Data Manipulation. R package version 1.0.5. https://CRAN.R-project.org/package=dplyr

51. Jerome Friedman, Trevor Hastie, Robert Tibshirani (2010). Regularization Paths for Generalized Linear Models via Coordinate Descent. Journal of Statistical Software, 33(1), 1-22. URL https://www.jstatsoft.org/v33/i01/.

52. Katharine M. Mullen and Ivo H. M. van Stokkum (2012). nnls: The Lawson-Hanson algorithm for non-negative least squares (NNLS). R package version 1.4. https://CRAN.R-project.org/package=nnls

53. Taiyun Wei and Viliam Simko (2021). R package "corrplot": Visualization of a Correlation Matrix (Version 0.89). Available from https://github.com/taiyun/corrplot

Figure 1. Data Import page for the binary data analysis section

Figure 2. Summary statistics page for the binary data analysis section

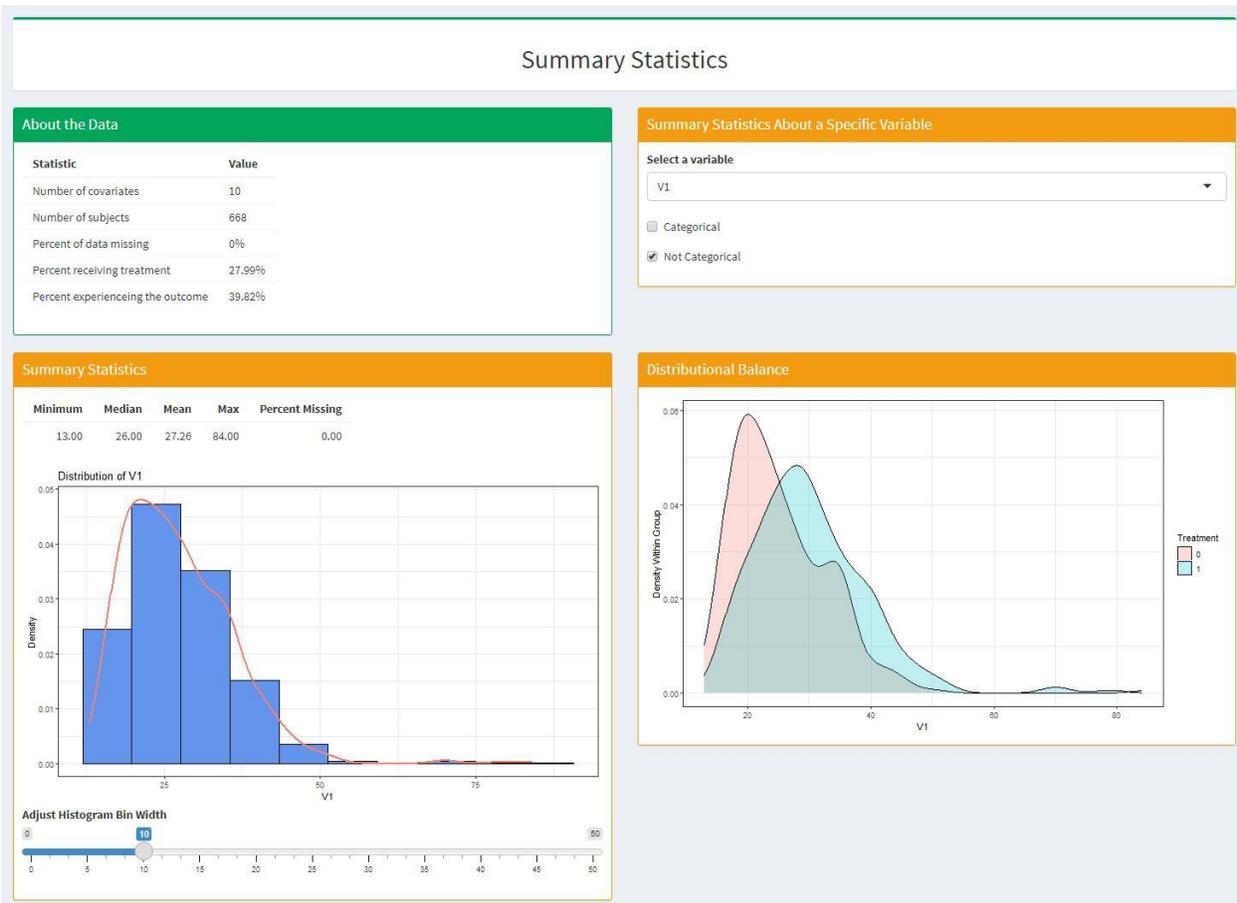

Figure 3. Summary statistics page for the binary data analysis section

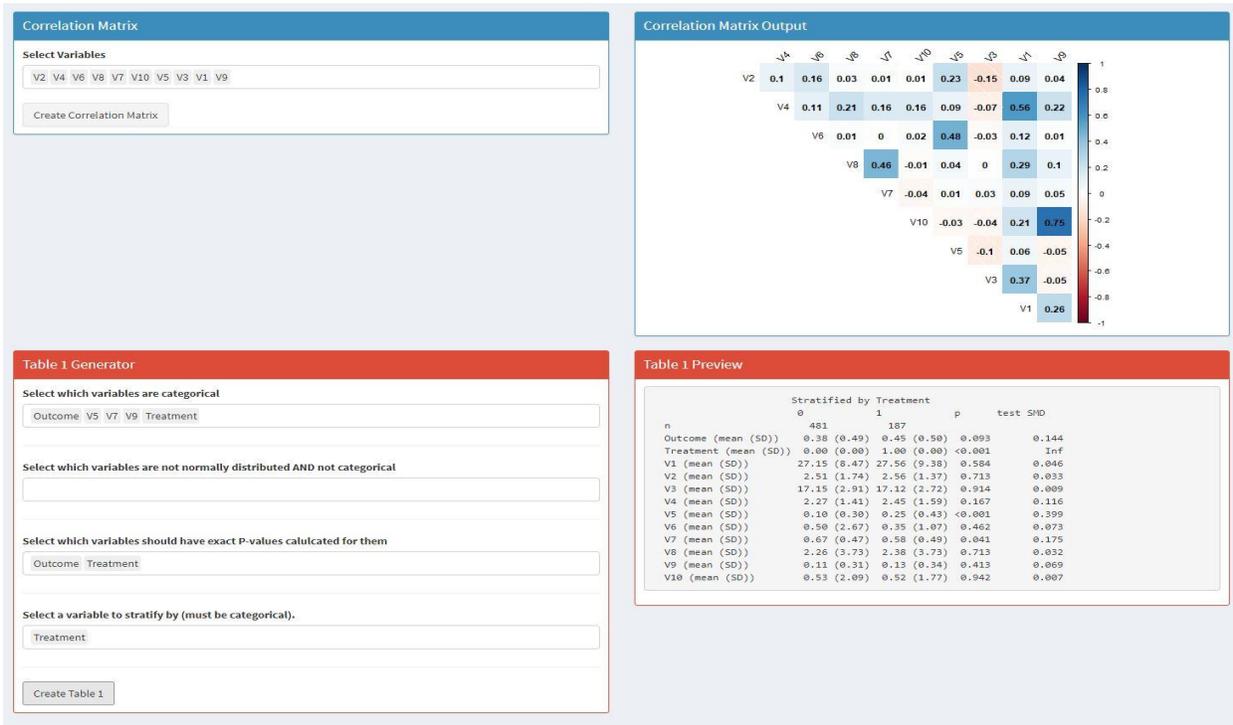

Figure 4. Results page for the binary data analysis section

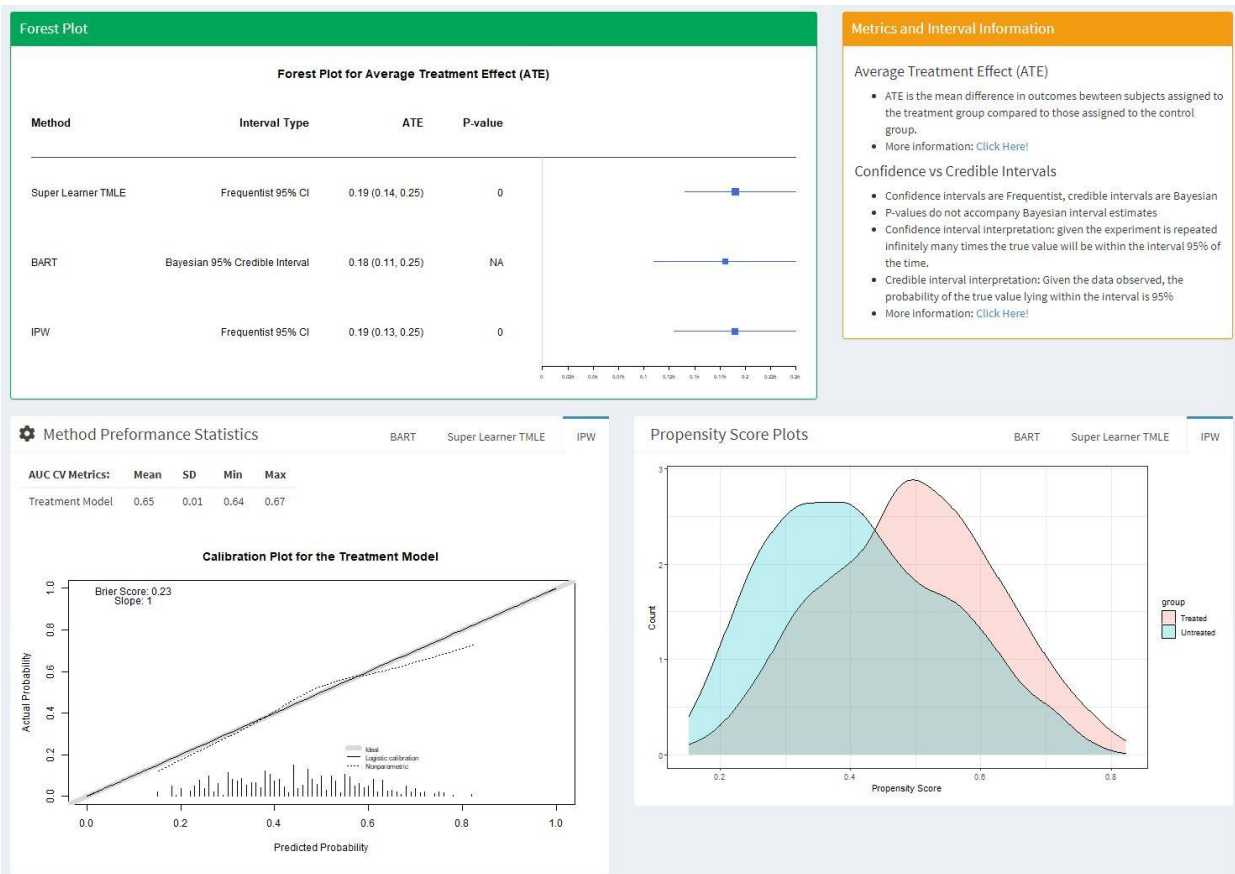

Figure 5. Data Import page for the continuous data analysis section

Figure 6. Summary statistics page for the continuous data analysis section

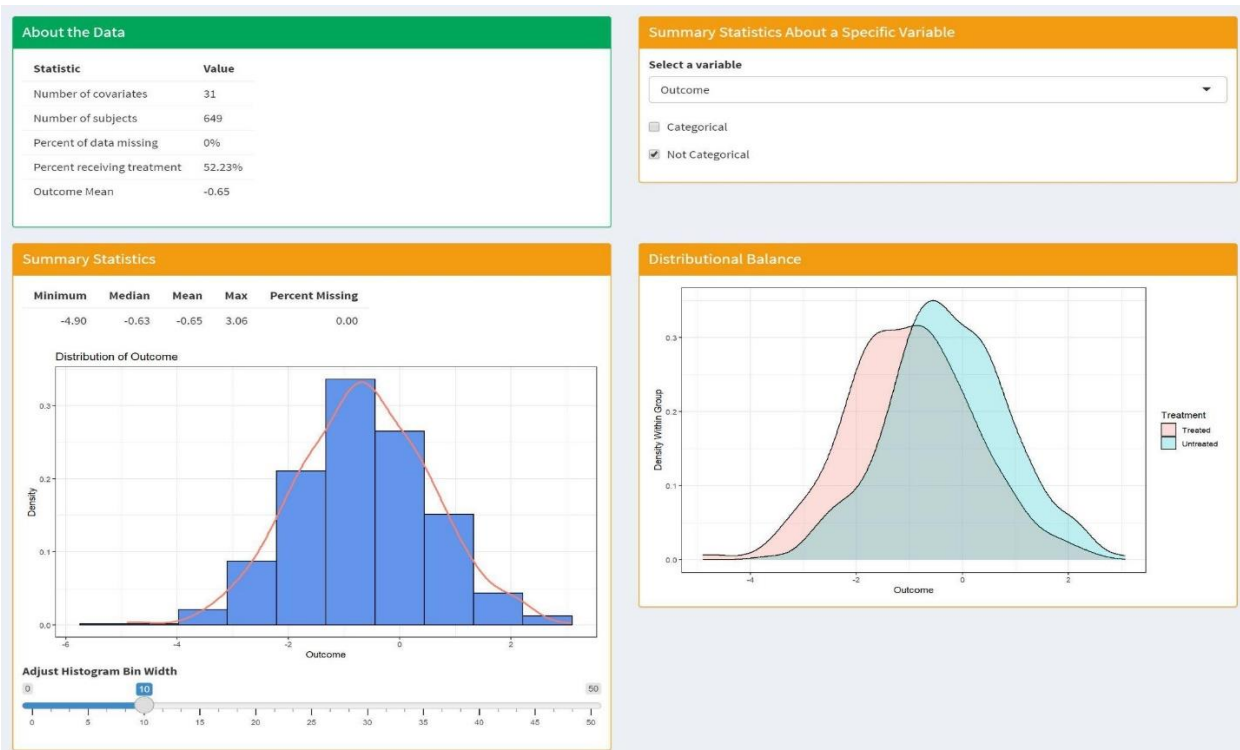

Figure 7. Summary statistics page for the continuous data analysis section

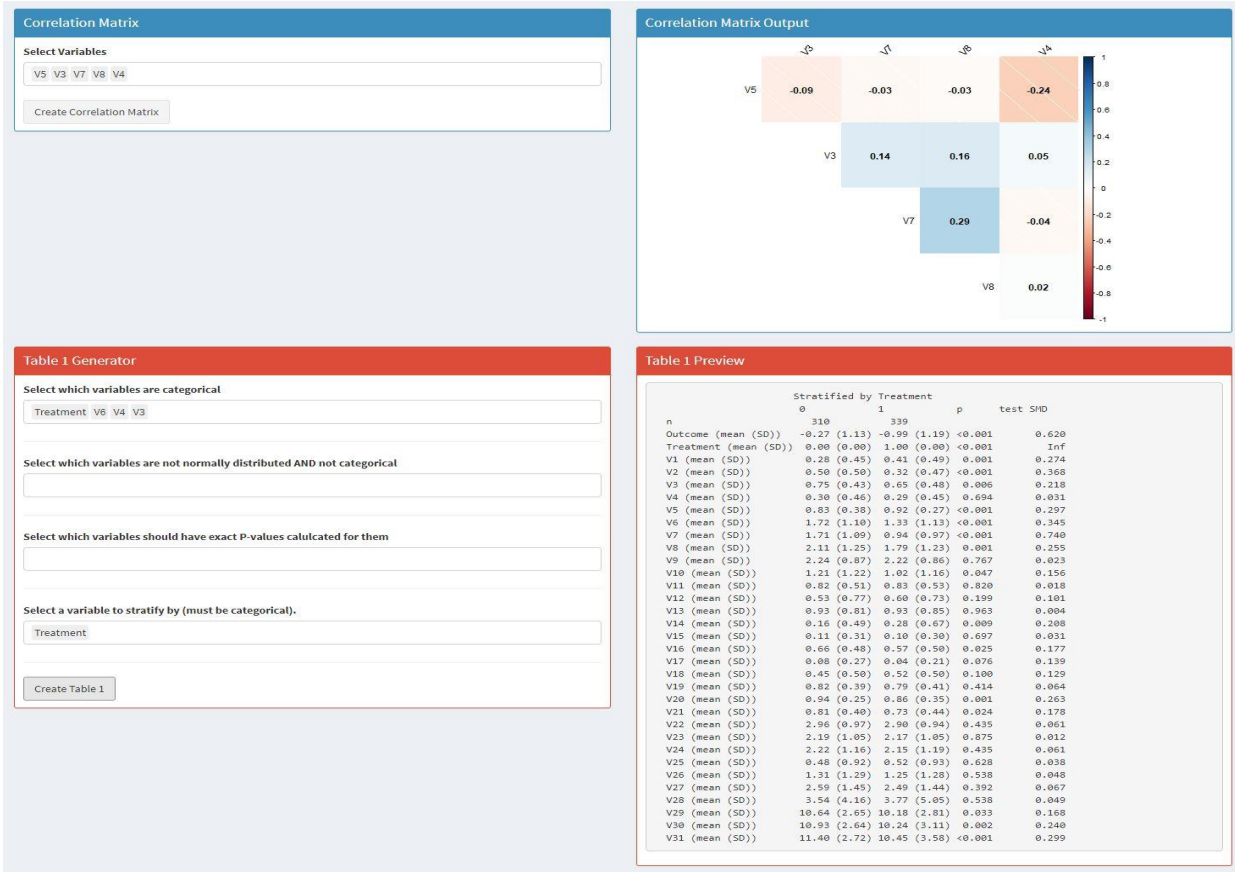

Figure 8. Results page for the continuous data analysis section

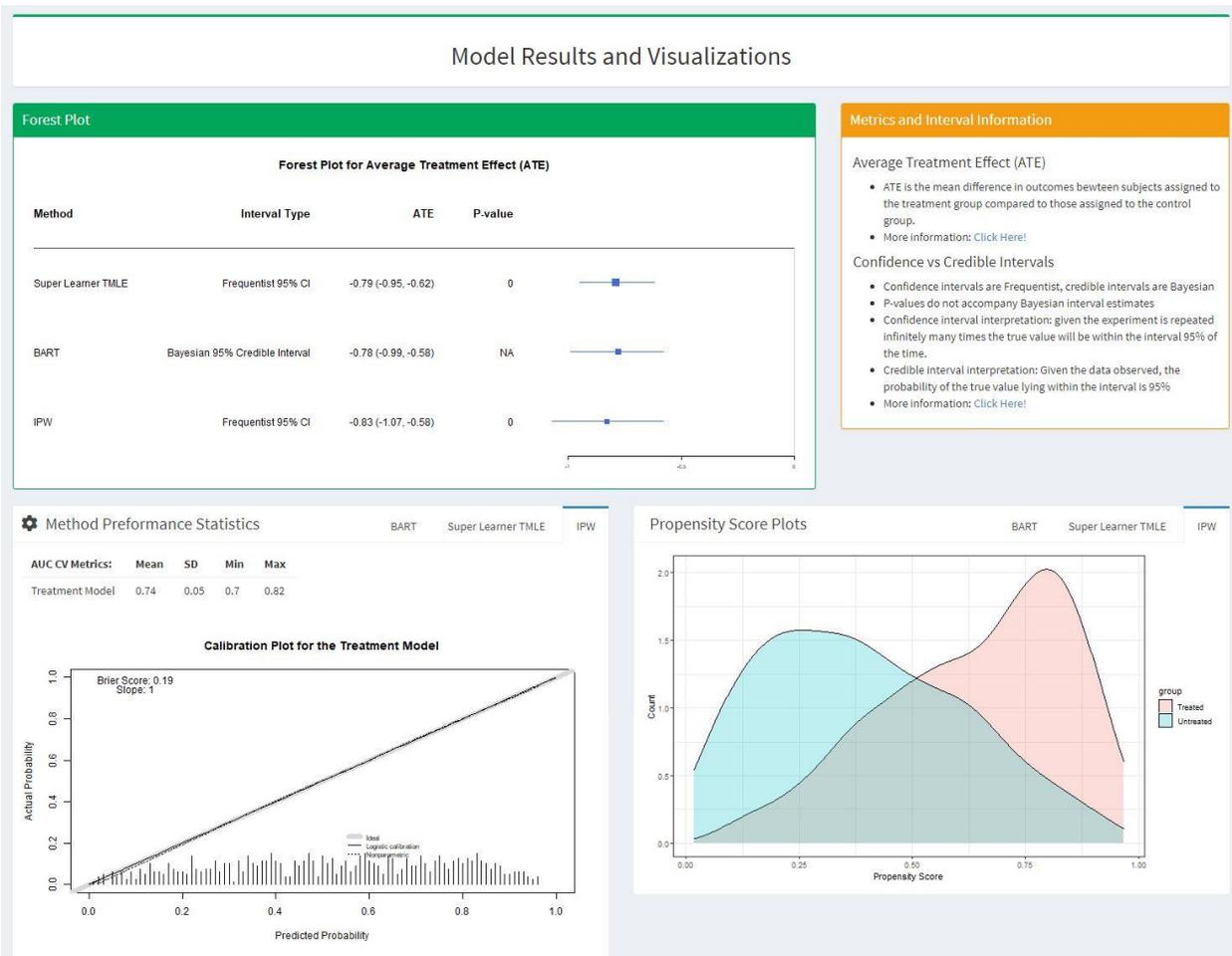

Figure 9. Data Import page for the survival data analysis section

Figure 10. Summary statistics page for the survival data analysis section

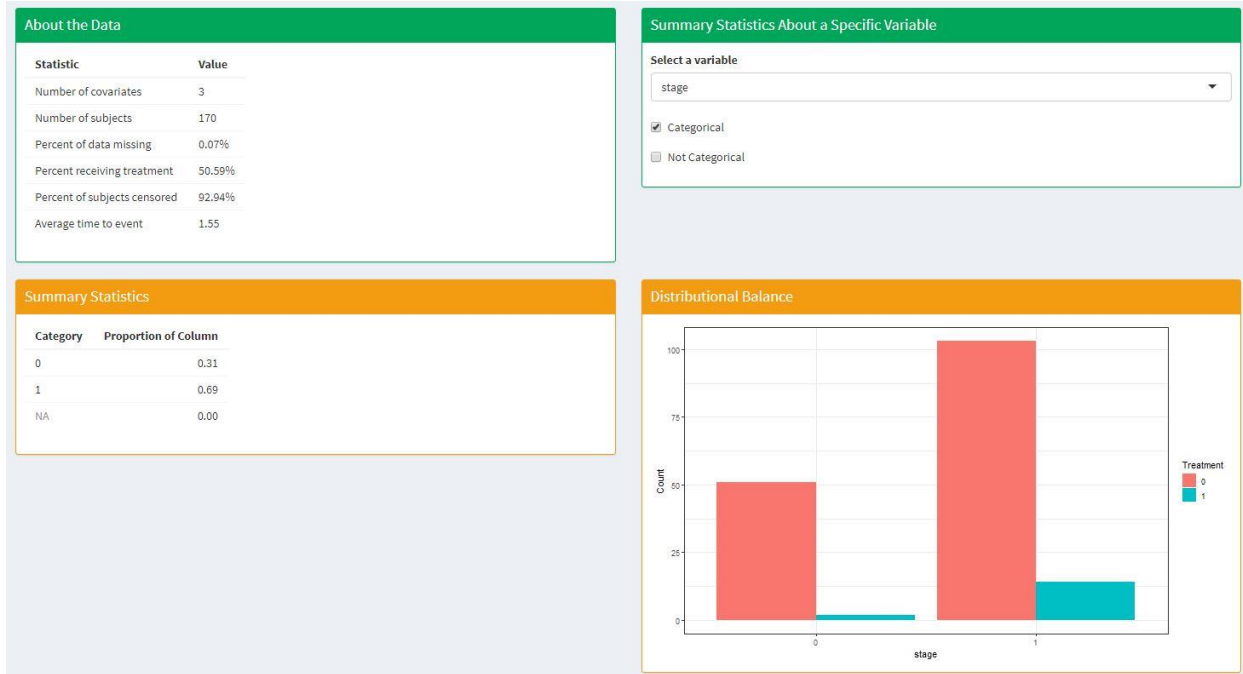

Figure 11. Summary statistics page for the survival data analysis section

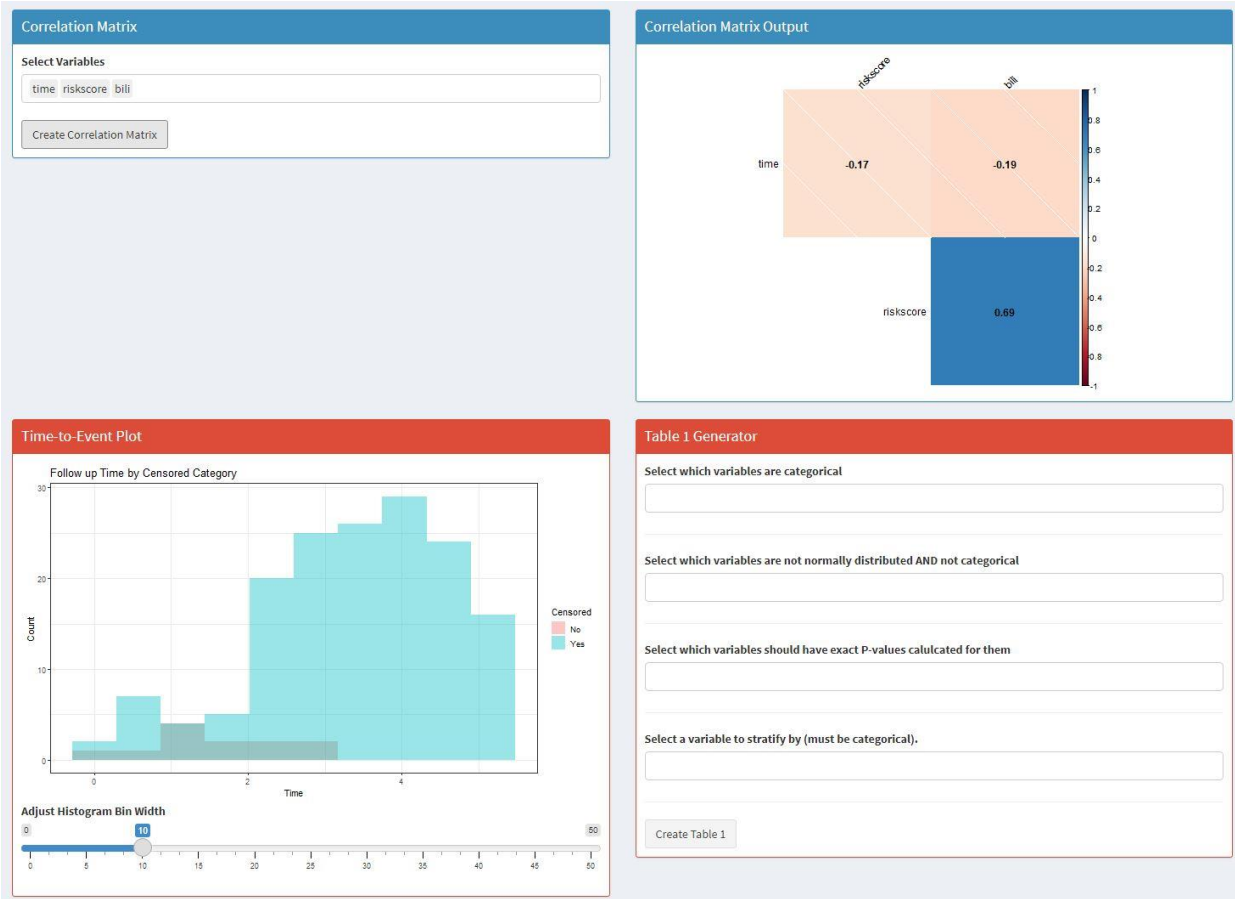

Figure 12. Results page for the survival data analysis section

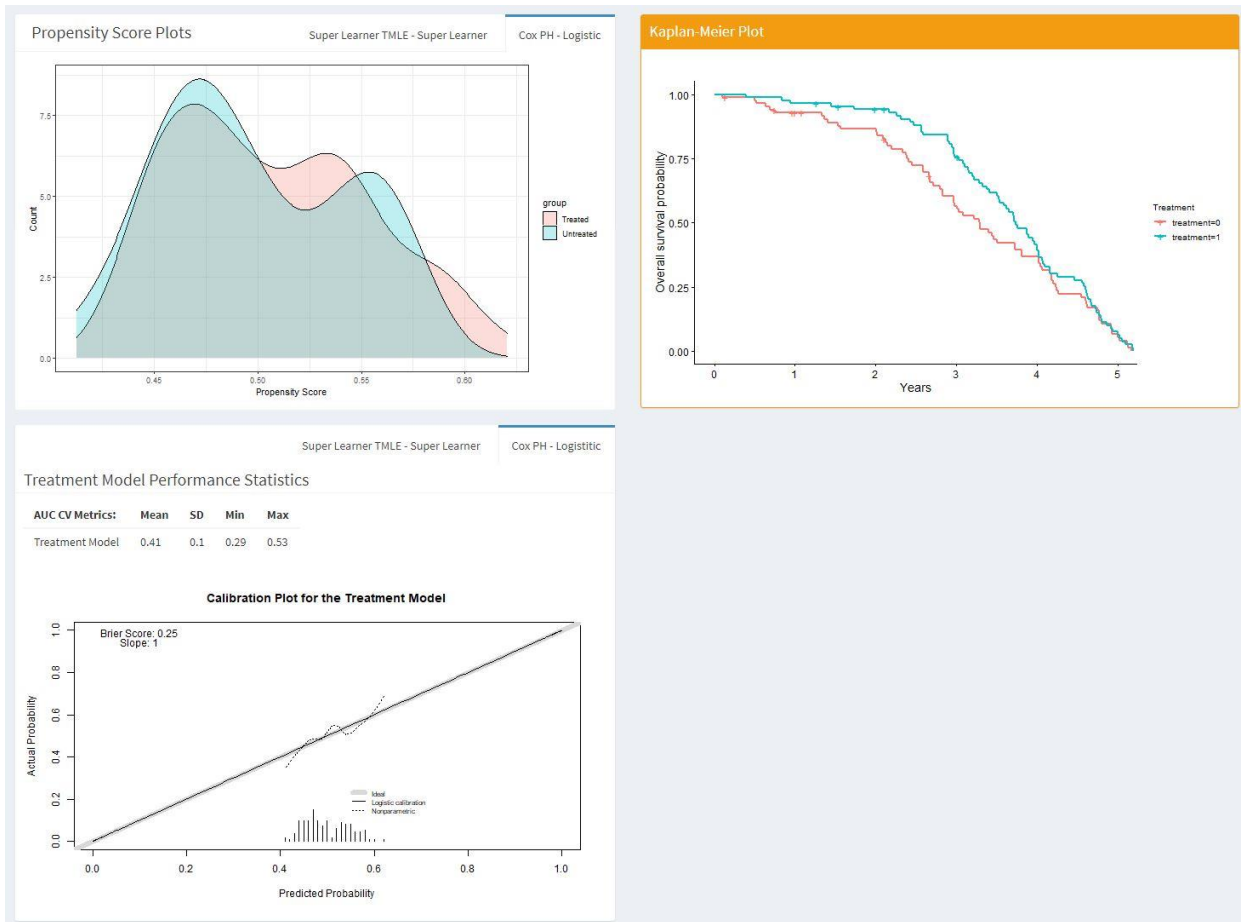

Figure 13. Results page for the survival data analysis section

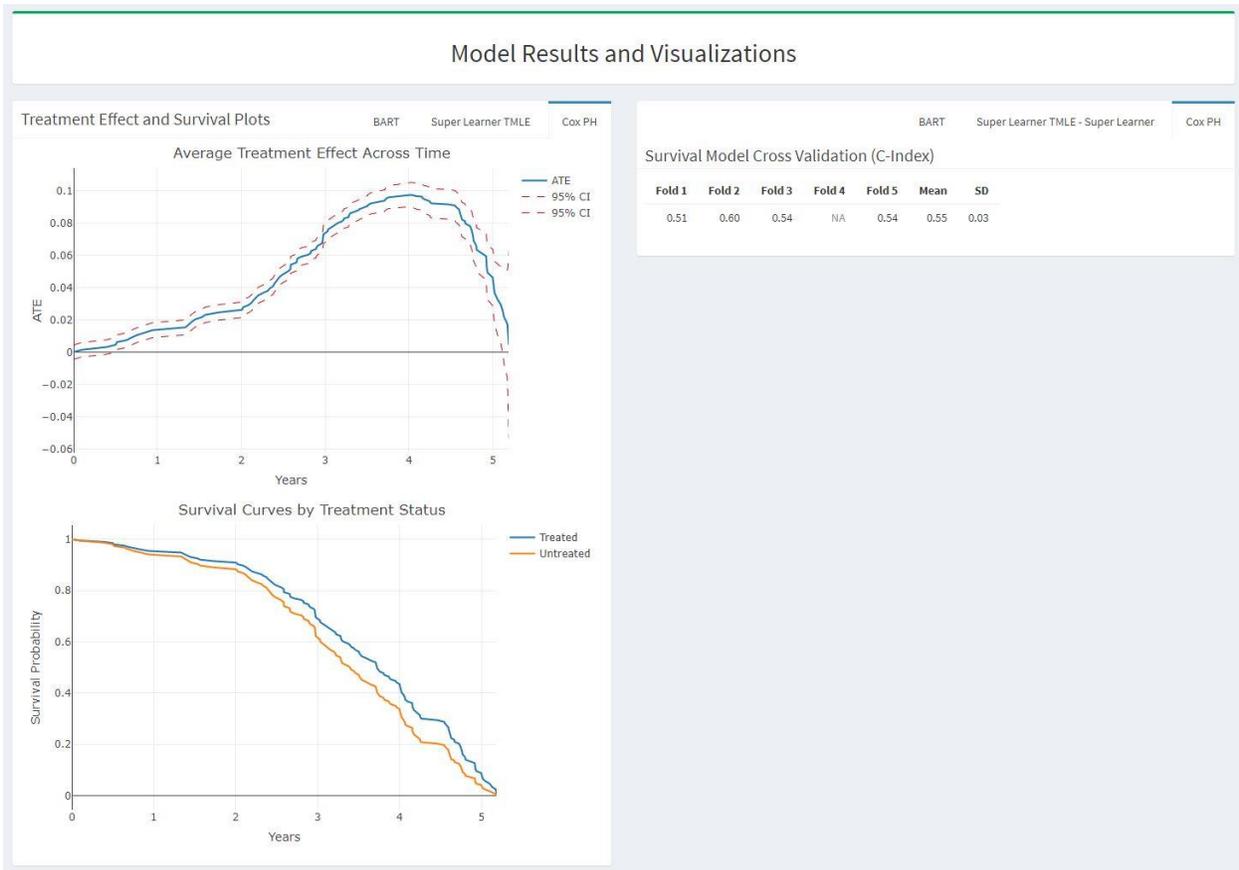

Figure 14. Back-end outline for the binary and continuous data analysis sections

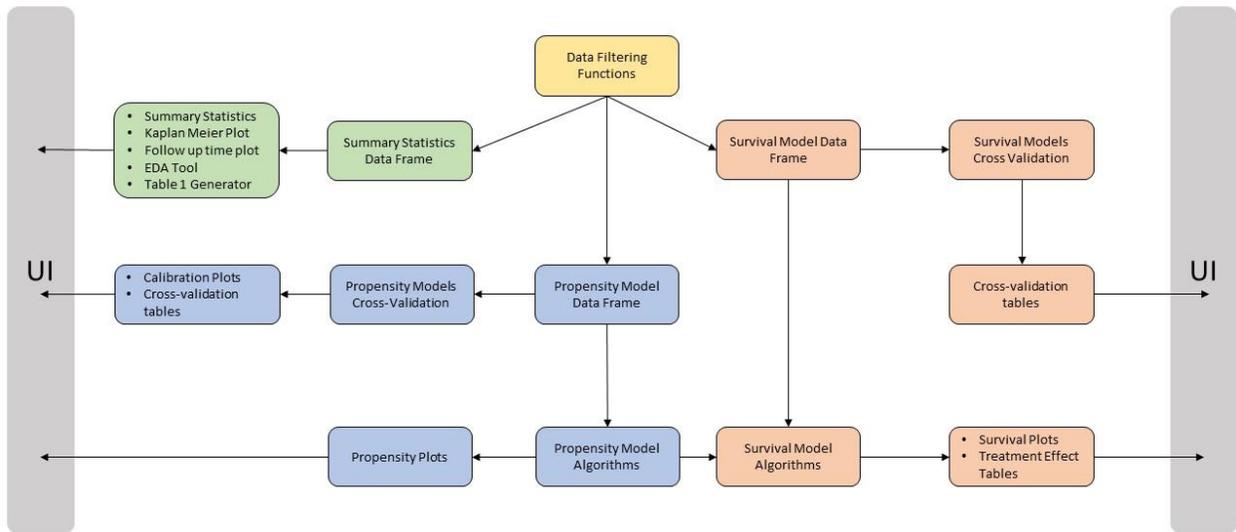

Figure 15. Back-end outline for the survival data analysis section

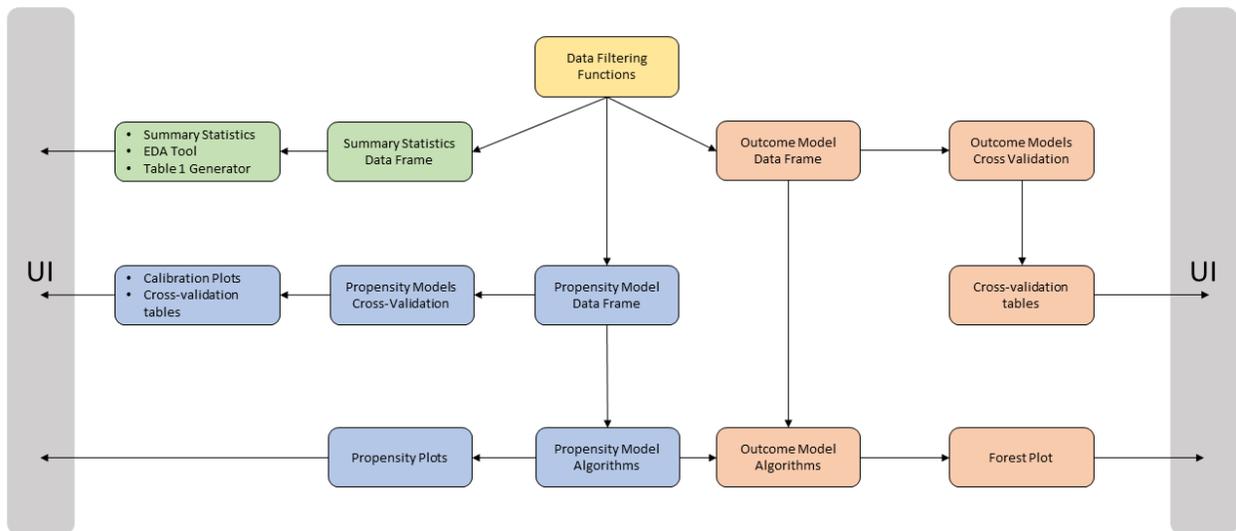